\documentclass{aa}  
\usepackage{natbib}
\bibpunct{(}{)}{;}{a}{}{,} 

\usepackage[T1]{fontenc}
\usepackage{graphicx}
\usepackage{amssymb}
\usepackage{subcaption}
\usepackage{cleveref}
\usepackage{booktabs}

\usepackage{epstopdf}
\usepackage{gensymb}
\usepackage{siunitx}
\usepackage[american]{circuitikz}
\usetikzlibrary{shapes,arrows}
\usepackage{mathtools}
\usepackage{soul}
\usepackage{commath}
\usepackage{afterpage}
\usepackage{placeins}
\usepackage{diagbox}
\usepackage{makecell}
\usepackage{tabularx}
\usepackage{cancel} 

\begin{document}

   \title{System equivalent flux density of Stokes I, Q, U, V of a polarimetric interferometer}


   \author{A. T. Sutinjo
          \inst{1}
           \and
           D. C. X. Ung
          \inst{1}
          \and
          M. Sokolowski
          \inst{1}
          }

   \institute{International Centre for Radio Astronomy Research (ICRAR), Curtin University, 6102 Australia\\
              \email{adrian.sutinjo@curtin.edu.au}
             }

   \date{
   Final \today
   }

 
  \abstract
  {} 
   {We present the system equivalent flux density (SEFD) expressions for all four Stokes parameters: I, Q, U, V.}
   {The expressions were derived based on our derivation of SEFD~I (for Stokes I) and subsequent extensions of that work to phased array and multipole interferometers. The key to the derivation of the SEFD~Q, U, V  expressions is to recognize that the noisy estimates of Q, U, V can be written as the trace of a matrix product. This shows that the SEFD~I is a special case, where the general case involves a diagonal or anti-diagonal $2\times2$ matrix interposed in the matrix multiplication. Following this step, the relation between the SEFD for I and Q, U, V becomes immediately evident. }
   {We present example calculations for a crossed dipole based on the formulas derived and the comparison between simulation and observation using the Murchison Widefield Array (MWA).}
   {}


   \maketitle
%
\section{Introduction}
\label{sec:intro}

System equivalent flux density (SEFD) is an important figure of merit of a radio telescope. Our previous work in~\citet{Sutinjo_AA2021} (Paper~I, dual-polarized antennas) and the extensions thereof~\citet{Sutinjo_2022} (Paper~II, tripole/multipole antennas) and \citet{Sutinjo_Array_AnA} (Paper~III, phased array) showed the derivation and validation of SEFD of Stokes I for a polarimetric interferometer. This paper expands the formula to all other Stokes parameters, Q, U, V. Recent discoveries in polarized astrophysical radio transients highlight the importance of knowing the telescope sensitivity for all four Stokes parameters over a wide field of view. A detailed review of the scientific motivation and application of this work is given in Sect.~\ref{sec_science_apps}. 

The expressions we shall derive are valid for dual-polarized antennas up to multipole antennas. The dual-polarized antenna systems are typical of ground-based polarimetric interferometer whereas tripole antennas are being proposed for a space-borne intereferometer~\citep{Chen_2018ExA....45..231C, Huang_2018, Bentum_2020AdSpR..65..856B, Chen_doi:10.1098/rsta.2019.0566}. Therefore, to illustrate this generality, we begin with a tripole antenna system as an example. The extension to a multipole system is straightforward; the simplification to the dual-polarized antenna system will be discussed in Sect.~\ref{sec:SEFD_dual}.  

The voltages measured by the tripole antenna system~1 in response to the target electric field ($|_t$) are~(see Sect. 1 of Paper~II)
\begin{eqnarray}
\mathbf{v}_1|_t&=&\mathbf{J}_1\mathbf{e}_t, \nonumber \\
\left[ \begin{array}{c}
V_{X1}|_t \\
V_{Y1}|_t \\
V_{Z1}|_t
\end{array} \right]
&=&
\left[ \begin{array}{cc}
l_{X1\theta} & l_{X1\phi} \\
l_{Y1\theta} & l_{Y1\phi} \\
l_{Z1\theta} & l_{Z1\phi} \\
\end{array} \right]
\left[ \begin{array}{c}
E_{t\theta} \\
E_{t\phi} 
\end{array} \right],
\label{eqn:J1}
\end{eqnarray} 
where $\mathbf{J}_1$ is the Jones matrix of system~1, with entries which represent the antenna effective lengths, in units of \si{\metre}~(see Paper~I); $E_{t\theta},E_{t\phi}$ are the components of the incident electric field vector, in units of \si{\volt\per\metre}. The voltages are correlated, resulting in the outer product 
\begin{eqnarray}
\mathbf{v}_1\mathbf{v}_2^H|_t=\mathbf{J}_1\mathbf{e}_t\mathbf{e}_t^H\mathbf{J}_2^H.
\label{eqn:outer_voltage}
\end{eqnarray}
where $\mathbf{J}_2$ is the Jones matrix of system 2, $.^H$ denotes the conjugate transpose. 

The quantities of interest are the linear combinations of the entries of $\mathbf{e}_t\mathbf{e}_t^H$ in Eq.~\eqref{eqn:outer_voltage}. The expected value is
\begin{eqnarray}
\left<\mathbf{e}_t\mathbf{e}_t^H\right>&=&
\left[ \begin{array}{cc}
\left<|E_{t\theta}|^2\right>& \left<E_{t\theta}E_{t\phi}^*\right> \\
\left<E_{t\theta}^*E_{t\phi} \right> & \left<|E_{t\phi}|^2\right>
\end{array} \right]\nonumber \\
&=&\frac{1}{2}\left[ \begin{array}{cc}
I+Q & U-jV \\
U+jV & I-Q
\end{array}\right]
\label{eqn:outer_field}
\end{eqnarray}
as discussed in Paper~I. Eq.~\eqref{eqn:outer_field} suggests that the corresponding random variables from which the Stokes parameters are inferred (we drop the subscript $._t$ for brevity)
\begin{eqnarray}
\tilde{I}&=&(\tilde{\mathbf{e}}\tilde{\mathbf{e}}^H)_{1,1}+(\tilde{\mathbf{e}}\tilde{\mathbf{e}}^H)_{2,2}, \nonumber \\
\tilde{Q}&=&(\tilde{\mathbf{e}}\tilde{\mathbf{e}}^H)_{1,1}-(\tilde{\mathbf{e}}\tilde{\mathbf{e}}^H)_{2,2}, \nonumber\\
\tilde{U}&=&(\tilde{\mathbf{e}}\tilde{\mathbf{e}}^H)_{1,2}+(\tilde{\mathbf{e}}\tilde{\mathbf{e}}^H)_{2,1}, \nonumber\\
\tilde{V}&=&j\left[(\tilde{\mathbf{e}}\tilde{\mathbf{e}}^H)_{1,2}-(\tilde{\mathbf{e}}\tilde{\mathbf{e}}^H)_{2,1}\right].
\label{eqn:Noisy_est}
\end{eqnarray}
The SEFD expressions are proportional to the standard deviations of Eq.~\eqref{eqn:Noisy_est}. Sect.~\ref{sec:method} discusses how to obtain them. 

\section{Method: SEFD derivation}
\label{sec:method}


\subsection{Expressing noisy estimates using matrix trace}
\label{sec:mat_trace}
 
To obtain $\tilde{\mathbf{e}}\tilde{\mathbf{e}}^H$ from $\mathbf{v}_1\mathbf{v}_2^H$, we need to remove the influence of the Jones matrices. This is possible by forming the left inverse of the Jones matrix, $\mathbf{L}=(\mathbf{J}^H\mathbf{J})^{-1}\mathbf{J}^H$~(see Sect.~2 of Paper~II). We note that for a dual-polarized system, the left inverse is identical to the matrix inverse. Of course, for the left inverse to exist, the columns of the Jones matrices must be linearly independent, which is not difficult to build into the antenna design.  

Using the left inverse, the result is an estimate~(see Sect.~2 of Paper II)
\begin{eqnarray}
\tilde{\mathbf{e}}\tilde{\mathbf{e}}^H&=&\left(\mathbf{J}_1^H\mathbf{J}_1\right)^{-1}\mathbf{J}_1^H\mathbf{v}_1\mathbf{v}_2^H\mathbf{J}_2\left(\mathbf{J}^H_2\mathbf{J}_2\right)^{-H}\nonumber \\
&=&\mathbf{L}_1\mathbf{v}_1\mathbf{v}_2^H\mathbf{L}_2^H.
\label{eqn:pol}
\end{eqnarray}
To obtain the noisy estimates of the Stokes parameters in Eq.~\eqref{eqn:Noisy_est}, we use the matrix trace operation, $\text{tr}()$, on Eq.~\eqref{eqn:pol}. As an example, we start with $\tilde{I}$. 
\begin{eqnarray}
\tilde{I}=\text{tr}\left(\tilde{\mathbf{e}}\tilde{\mathbf{e}}^H\right)
=\text{tr}\left(\mathbf{L}_1\mathbf{v}_1\mathbf{v}_2^H\mathbf{L}_2^H\right).
\label{eqn:I_tilde}
\end{eqnarray}
Applying the trace property of matrix product\footnote{This can be shown by exchanging the summation order, for example see planetmath.org/proofofpropertiesoftraceofamatrix, www.statlect.com/matrix-algebra/trace-of-a-matrix, en.wikipedia.org/wiki/Trace\_(linear\_algebra)\#cite\_{note-4}}, $\text{tr}(AB)=\text{tr}(BA)$ which is valid for $A~(m\times n)$ and $B~(n\times m$), successively to Eq.~\eqref{eqn:I_tilde}, leads to 
\begin{eqnarray}
\tilde{I}
&=&\text{tr}\left(\mathbf{v}_1\mathbf{v}_2^H\mathbf{L}_2^H\mathbf{L}_1\right) =\text{tr}\left(\mathbf{v}_2^H\mathbf{L}_2^H\mathbf{L}_1\mathbf{v}_1\right) \nonumber \\
&=&\mathbf{v}_2^H\mathbf{L}_2^H\mathbf{L}_1\mathbf{v}_1.
\label{eqn:I_tilde2}
\end{eqnarray}
The last equality  in Eq.~\eqref{eqn:I_tilde2} comes from the fact that $\mathbf{v}_2^H\mathbf{L}_2^H\mathbf{L}_1\mathbf{v}_1$ produces a single number such that the trace operator is redundant. Similarly, for $\tilde{Q}$,
\begin{eqnarray}
\tilde{Q}&=&\text{tr}\left(\mathbf{P}_Q\tilde{\mathbf{e}}\tilde{\mathbf{e}}^H\right)
=\text{tr}\left(\mathbf{P}_Q\mathbf{L}_1\mathbf{v}_1\mathbf{v}_2^H\mathbf{L}_2^H\right) \nonumber \\
&=&\mathbf{v}_2^H\mathbf{L}_2^H\mathbf{P}_Q\mathbf{L}_1\mathbf{v}_1,
\label{eqn:Q_tilde}
\end{eqnarray}
where 
\begin{eqnarray}
\mathbf{P}_Q=
\left[ \begin{array}{cc}
1 & 0 \\
0 & -1 \\
\end{array} \right].
\label{eqn:PQ}
\end{eqnarray}
Similarly, 
\begin{eqnarray}
\tilde{U}
&=&\mathbf{v}_2^H\mathbf{L}_2^H\mathbf{P}_U\mathbf{L}_1\mathbf{v}_1,
\label{eqn:U_tilde}  \\
\tilde{V}
&=&\mathbf{v}_2^H\mathbf{L}_2^H\mathbf{P}_V\mathbf{L}_1\mathbf{v}_1,
\label{eqn:V_tilde}
\end{eqnarray}
where  
\begin{eqnarray}
\mathbf{P}_U&=&
\left[ \begin{array}{cc}
0 & 1 \\
1 & 0 \\
\end{array} \right],
\label{eqn:PU} \\
\mathbf{P}_V&=&
j\left[ \begin{array}{cc}
0 & -1 \\
1 & 0 \\
\end{array} \right].
\label{eqn:PV}
\end{eqnarray}
We now see that $\tilde{I}$ in Eq.~\eqref{eqn:I_tilde2} is a special case in which $\mathbf{P}_I=\mathbf{I}$, that is the identity matrix. For brevity we will refer to $\mathbf{L}_2^H\mathbf{P}_S\mathbf{L}_1=\mathbf{M}_S$ where the subscript $._S$ denotes the Stokes parameter of interest: $I, Q, U$, or $V$.

\subsection{SEFD formula}
\label{sec:SEFD_Fmla}
Following the same derivation steps as~(see Sect.~2, Sect.~3, Appendix of Paper~II) we can show that
\begin{eqnarray}
\text{SEFD}_{S} = \frac{4k}{ \eta_0}\sqrt{\mathbf{t}_{\rm sysR1}^T\left(\mathbf{M}_{S}\circ \mathbf{M}_{S}^*\right)\mathbf{t}_{\rm sysR2}}, 
\label{eqn:SEFD_gen}
\end{eqnarray}
where $\circ$ refers to the element-by-element product, $.^*$ denotes complex conjugation such that $\mathbf{M}_{S}\circ \mathbf{M}_{S}^*$ produces the magnitude squared of each entry of $\mathbf{M}_{S}$; 
\begin{eqnarray}
\mathbf{M}_{S=I,Q,U,V}=\mathbf{L}_2^H\mathbf{P}_{S=I,Q,U,V}\mathbf{L}_1,
\label{eqn:MS}
\end{eqnarray}
where $\mathbf{P}_I, \mathbf{P}_Q, \mathbf{P}_U, \mathbf{P}_V$ are shown in Sect~\ref{sec:mat_trace} and
\begin{eqnarray}
\mathbf{t}_{\rm sysR1}=\left[ \begin{array}{c}
T_{{\rm sys}X1}R_{{\rm ant}, X1}\\
T_{{\rm sys}Y1}R_{{\rm ant}, Y1} \\
T_{{\rm sys}Z1}R_{{\rm ant}, Z1}
\end{array} \right],
\label{eqn:TR1}
\end{eqnarray}
and similarly for $\mathbf{t}_{\rm sysR2}$ (with entries that pertain to $T_{sys}$'s and $R_{ant}$'s of system 2).

\subsection{Special case: dual-polarized antennas with an identical antenna design}
\label{sec:SEFD_dual}
We consider a typical case of X and Y dual-polarized antennas which are common in ground-based radio astronomy. In this case, the last row in Eq.~\eqref{eqn:J1} is removed. Furthermore, an identical antenna design implies an identical Jones matrix, $\mathbf{J}_1=\mathbf{J}_2=\mathbf{J}_{dual}$, where
\begin{eqnarray}
\mathbf{J}_{dual}=
\left[ \begin{array}{cc}
l_{X\theta} & l_{X\phi} \\
l_{Y\theta} & l_{Y\phi} \\
\end{array} \right].
\label{eqn:Jdual}
\end{eqnarray} 
The left inverse becomes the inverse, $\mathbf{L}=\mathbf{J}_{dual}^{-1}$ such that 
\begin{eqnarray}
\mathbf{M}_{S:dual}=\mathbf{J}_{dual}^{-H}\mathbf{P}_{S}\mathbf{J}_{dual}^{-1}.
\label{eqn:Mdual}
\end{eqnarray}
As a result
\begin{eqnarray}
\text{SEFD}_{S:dual} = \frac{4k}{ \eta_0}\sqrt{\mathbf{t}_{\rm sysR1}^T\left(\mathbf{M}_{S:dual}\circ \mathbf{M}_{S:dual}^*\right)\mathbf{t}_{\rm sysR2}}. 
\label{eqn:SEFD_gen2}
\end{eqnarray}

It can be shown that (see Sect~2.5 of Paper~III)
\begin{eqnarray}
\mathbf{M}_{I:dual}=\frac{1}{\norm{\mathbf{l}_X-\mathbf{p}}^2\norm{\mathbf{l}_{Y}}^2}\left[ \begin{array}{cc}
\norm{\mathbf{l}_{Y}}^2 & -\mathbf{l}_X^{H}\mathbf{l}_Y \\
-\mathbf{l}_X^{H}\mathbf{l}_Y & \norm{\mathbf{l}_{X}}^2
\end{array} \right],
\label{eqn:MIdual}
\end{eqnarray}
where $\norm{.}$ operator refers to the length of the vector in question, such that $\norm{\mathbf{l}_X}$ and $\norm{\mathbf{l}_Y}$ are the lengths of the row vectors of $\mathbf{J}_{dual}$, $\mathbf{l}_X=[l_{X\theta}, l_{X\phi}]^T$, similarly with $\mathbf{l}_Y$. The denominator in the right hand side of Eq.~\eqref{eqn:MIdual} is the absolute value squared of the determinant of the Jones matrix.
\begin{eqnarray}
|D|^2=|l_{X\theta}l_{Y\phi}-l_{X\phi}l_{Y\theta}|^2
=\norm{\mathbf{l}_X-\mathbf{p}}^2\norm{\mathbf{l}_{Y}}^2, 
\label{eqn:det}
\end{eqnarray}
where $\mathbf{p}$ is the projection vector of $\mathbf{l}_X$ onto $\mathbf{l}_Y$. By expanding the matrix algebra, we find for Stokes~Q,
\begin{eqnarray}
|D|^2\mathbf{M}_{Q}=\left[ \begin{array}{cc}
 |l_{Y\phi}|^2-|l_{Y\theta}|^2 & -l_{Y\phi}^*l_{X\phi}+l_{Y\theta}^*l_{X\theta}\\
 -l_{X\phi}^*l_{Y\phi}+l_{X\theta}^*l_{Y\theta}&  |l_{X\phi}|^2-|l_{X\theta}|^2
\end{array} \right],
\label{eqn:MQdual}
\end{eqnarray}
where we dropped the $._{:dual}$ subscript for brevity. For Stokes~U,
\begin{eqnarray}
|D|^2\mathbf{M}_{U}=\left[ \begin{array}{cc}
 -2\Re(l_{Y\phi}^*l_{Y\theta})
 & l_{Y\phi}^*l_{X\theta}+l_{Y\theta}^*l_{X\phi}\\
 l_{X\phi}^*l_{Y\theta}+l_{X\theta}^*l_{Y\phi}&   -2\Re(l_{X\phi}^*l_{X\theta})
\end{array} \right],
\label{eqn:MUdual} 
\end{eqnarray}
Finally for Stokes~V,
\begin{eqnarray}
\frac{|D|^2}{j}\mathbf{M}_{V}=\left[ \begin{array}{cc}
 2j\Im(l_{Y\phi}^*l_{Y\theta}) 
 & -l_{Y\phi}^*l_{X\theta}+l_{Y\theta}^*l_{X\phi}\\
 -l_{X\phi}^*l_{Y\theta}+l_{X\theta}^*l_{Y\phi}&   2j\Im(l_{X\phi}^*l_{X\theta})
\end{array} \right].
\label{eqn:MVdual}
\end{eqnarray}
Example results for a dual polarized dipoles and the Murchison Widefield Array (MWA) are shown next.

\section{Results}
\label{sec:res}

\subsection{X-Y short dipoles}
\label{sec:short_dip}
We begin by considering simple X-Y short dipoles. In this case, the Jones matrix is real
\begin{eqnarray}
\mathbf{J}
&=&\left[ \begin{array}{cc}
\cos\theta\cos\phi & -\sin\phi \\
\cos\theta\sin\phi & \cos\phi
\end{array} \right]\nonumber \\
&=&\left[ \begin{array}{cc}
\cos\phi & -\sin\phi \\
\sin\phi & \cos\phi
\end{array} \right]\left[ \begin{array}{cc}
\cos\theta  & 0 \\
0 & 1
\end{array} \right] 
=\mathbf{R}\mathbf{\Sigma}.
\label{eqn:Jdipole}
\end{eqnarray}
In Eq.~\eqref{eqn:Jdipole}, the antenna length in units of meter could be included by pre-multiplying the matrix with a scalar $l$. The decomposition in Eq.~\eqref{eqn:Jdipole} was discussed in Paper~I, where $\mathbf{R}$ is an orthogonal (rotation) matrix ($\mathbf{R}^T\mathbf{R}=\mathbf{R}\mathbf{R}^T=\mathbf{I}$, the identity matrix) and  $\mathbf{\Sigma}$ is a diagonal matrix that contains the singular values. As a result, we can write the following: $\mathbf{J}^{-1}=(\mathbf{R}\mathbf{\Sigma})^{-1}=\mathbf{\Sigma}^{-1}\mathbf{R}^{-1}=\mathbf{\Sigma}^{-1}\mathbf{R}^{T}$; $\mathbf{J}^{-H}=\mathbf{R}\mathbf{\Sigma}^{-1}$; and
\begin{eqnarray}
\mathbf{M}_{S}=\mathbf{J}^{-H}\mathbf{P}_S\mathbf{J}^{-1} = \mathbf{R}\mathbf{\Sigma}^{-1}\mathbf{P}_S\mathbf{\Sigma}^{-1}\mathbf{R}^{T}.
\label{eqn:M_S_ident}
\end{eqnarray}

Assuming identical antennas ($R_{antX}=R_{antY}=R_{ant}$) and the same system temperatures for X and Y ($T_{sysX}=T_{sysY}=T_{sys}$), it can be shown that Eq.~\eqref{eqn:Mdual} becomes~(see Paper~II)
\begin{eqnarray}
\text{SEFD}_{S} &=& \frac{4kR_{ant}T_{sys}}{ \eta_0}\sqrt{[1, 1]\left(\mathbf{M}_{S}\circ \mathbf{M}_{S}^*\right)[1, 1]^T} \nonumber\\
&=& \frac{4kR_{ant}T_{sys}}{ \eta_0}\sqrt{\text{tr}\left(\mathbf{M}_{S} \mathbf{M}_{S}^H\right)}.
\label{eqn:SEFD_dip}
\end{eqnarray}
Using Eq.~\eqref{eqn:M_S_ident}, 
\begin{eqnarray}
\mathbf{M}_{S} \mathbf{M}_{S}^H&=&\mathbf{R}\mathbf{\Sigma}^{-1}\mathbf{P}_S\mathbf{\Sigma}^{-1}\mathbf{R}^{T}\mathbf{R}\mathbf{\Sigma}^{-1}\mathbf{P}_S^{H}\mathbf{\Sigma}^{-1}\mathbf{R}^{T} \nonumber \\
&=& \mathbf{R}\mathbf{\Sigma}^{-1}\mathbf{P}_S\mathbf{\Sigma}^{-2}\mathbf{P}_S^{H}\mathbf{\Sigma}^{-1}\mathbf{R}^{T}. 
\label{eqn:MS_MSH}
\end{eqnarray}
The inverse of $\mathbf{\Sigma}$ in Eq.~\eqref{eqn:Jdipole} is easily done by inverting the diagonal entries. Using the relevant $\mathbf{P}_S$ matrix shown in Sect~\ref{sec:mat_trace}, we can show that
\begin{eqnarray}
\mathbf{M}_{I} \mathbf{M}_{I}^H=\mathbf{M}_{Q} \mathbf{M}_{Q}^H&=&\mathbf{R}\mathbf{\Sigma}^{-4}\mathbf{R}^{T} \nonumber\\
&=&\mathbf{R}\left[ \begin{array}{cc}
\cos^{-4}\theta  & 0 \\
0 & 1
\end{array} \right]\mathbf{R}^T.
\label{eqn:MI_MQ}
\end{eqnarray}
Since $\mathbf{R}$ is orthogonal, Eq.~\eqref{eqn:MI_MQ} is an eigendecomposition of $\mathbf{M}_{I} \mathbf{M}_{I}^H$ and $\mathbf{M}_{Q} \mathbf{M}_{Q}^H$  where $\mathbf{\Sigma}^{-4}$ is the eigenvalue matrix. As a result, the trace of that matrix is simply the sum of the eigenvalues~\cite[see. Ch.~6]{Strang_ILA2016}; the same result could also have been obtained by applying the trace of matrix product property shown in Sect~\ref{sec:mat_trace}, $\text{tr}(\mathbf{R}\mathbf{\Sigma}^{-4}\mathbf{R}^{T})=\text{tr}(\mathbf{\Sigma}^{-4}\mathbf{R}^{T}\mathbf{R})=\text{tr}(\mathbf{\Sigma}^{-4})$.
\begin{eqnarray}
\text{SEFD}_{I} =  \text{SEFD}_{Q}=
\frac{4kR_{ant}T_{sys}}{ \eta_0}\sqrt{\cos^{-4}\theta + 1}.
\label{eqn:SEFD_I_Q}
\end{eqnarray}
The $\text{SEFD}_{I}$ dependence as a function $\theta$ is the same as that identified in Sect.~4 of Paper~I. The effect of the metallic ground screen and/or actual antenna length can be easily added, if desired, as a multiplicative factor, $l(\theta)^{-2}$, in units of \si{\per\metre\squared} in the SEFD expression.

Similarly, by using Eq.~\eqref{eqn:MS_MSH} and substituting the relevant $\mathbf{P}_S$, we can show that
\begin{eqnarray}
\mathbf{M}_{U} \mathbf{M}_{U}^H=\mathbf{M}_{V} \mathbf{M}_{V}^H=\mathbf{R}\left(\mathbf{I}\cos^{-2}\theta\right)\mathbf{R}^{T}, 
\label{eqn:MU_MV}
\end{eqnarray}
such that 
\begin{eqnarray}
\text{SEFD}_{U} =  \text{SEFD}_{V}=
\frac{4kR_{ant}T_{sys}}{ \eta_0}\frac{\sqrt{2}}{\cos\theta }.
\label{eqn:SEFD_U_V}
\end{eqnarray}

\begin{figure}[htb]
\begin{center}
\noindent
\includegraphics[width=0.35\textwidth]{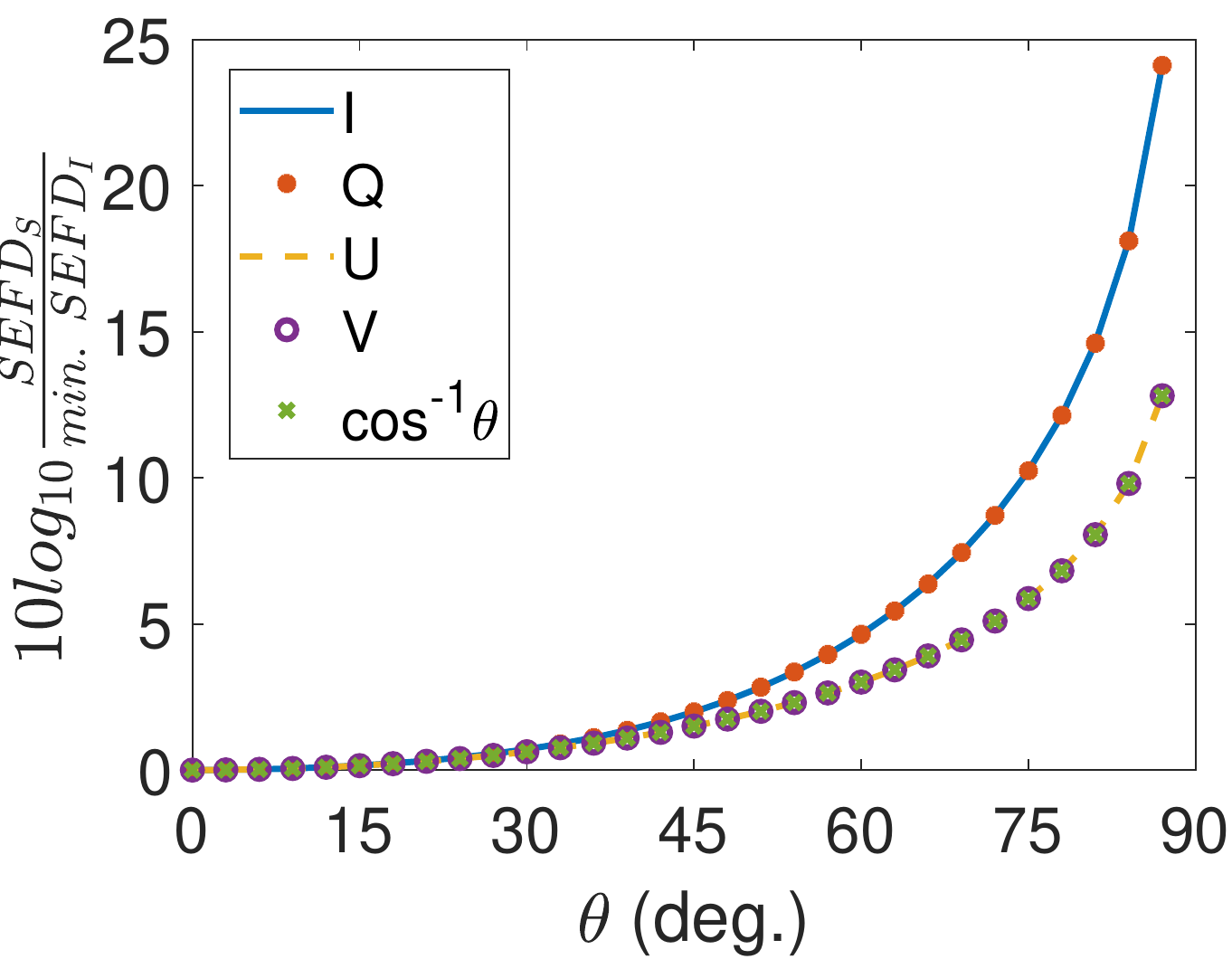}
\caption{$\text{SEFD}_{I,Q,U,V}/\text{min}.~\text{SEFD}_{I}$ in decibel (dB) as a function of zenith angle, $\theta$.}
\label{fig:SEFD_comp_dip}
\end{center}
\end{figure}

Therefore, for the orthogonal short dipole system, the SEFD is only dependent on $\theta$, the zenith angle. This was validated by numerical calculations based on Eq.~\eqref{eqn:SEFD_gen2} and the Jones matrix as originally presented in Eq.~\eqref{eqn:Jdipole}. 
Fig.~\ref{fig:SEFD_comp_dip} shows the SEFD~I, Q, U, V normalized to the minimum of $\text{SEFD}_I$ (at $\theta=0\degree$) as a function of zenith angle. It clearly shows that $\text{SEFD}_U$ and $\text{SEFD}_V$ rise more slowly than $\text{SEFD}_I$ and $\text{SEFD}_Q$ with increasing $\theta$ as expected from Eq.~\eqref{eqn:SEFD_I_Q} and Eq.~\eqref{eqn:SEFD_U_V}.


\subsection{MWA telescope pointing at $ZA = 46.1\degree,~Az = 80.5\degree$}
\label{sec:MWA_observation}

 \begin{figure*}[h]
  \centering
     \begin{subfigure}[t]{0.45\textwidth}
         \centering
          \includegraphics[width = \textwidth]{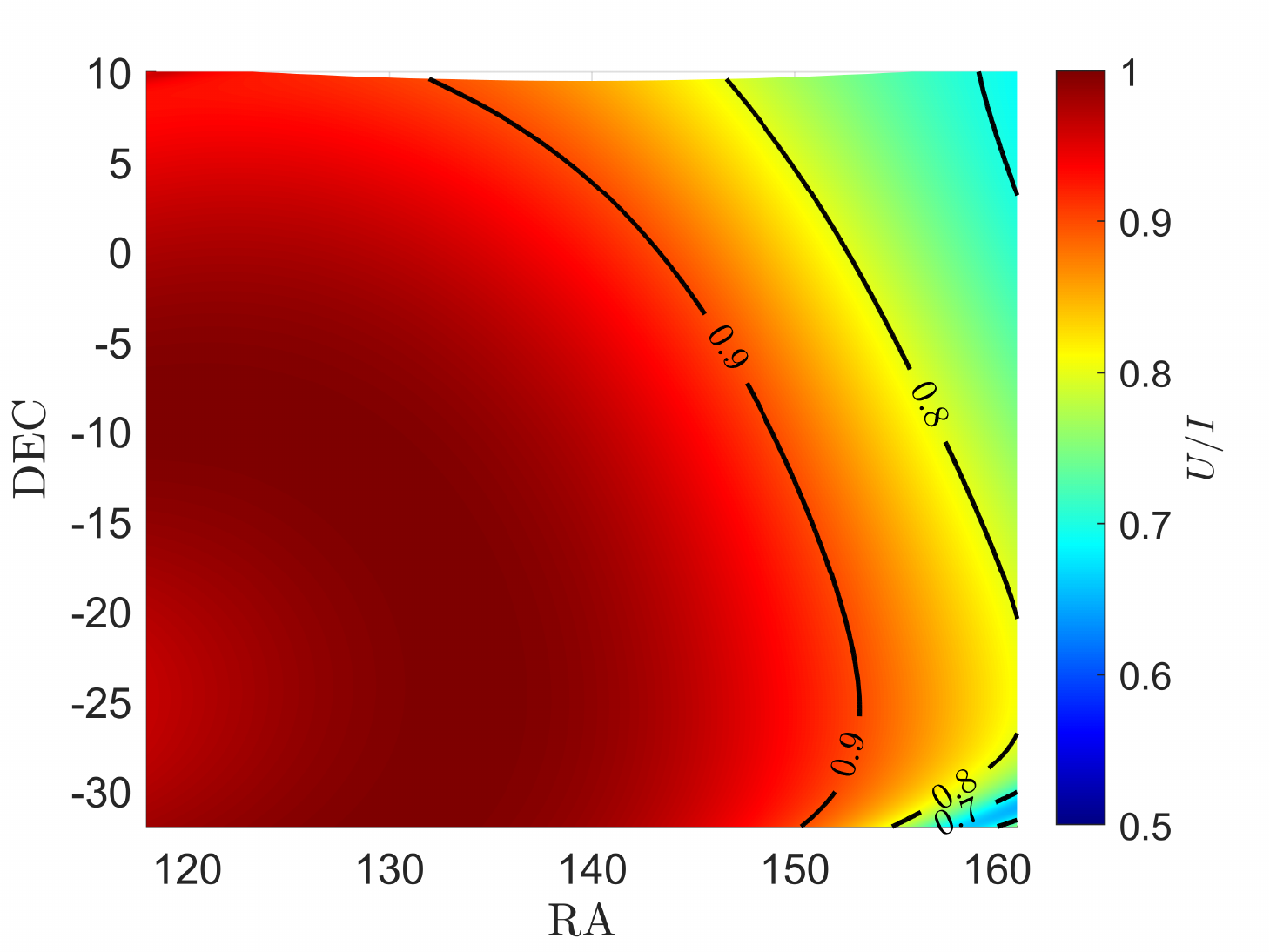}
         \caption{}
         \label{fig:sim_SEFDUI}
     \end{subfigure}
          \hfill
        \begin{subfigure}[t]{0.45\textwidth}
         \centering
          \includegraphics[width = \textwidth]{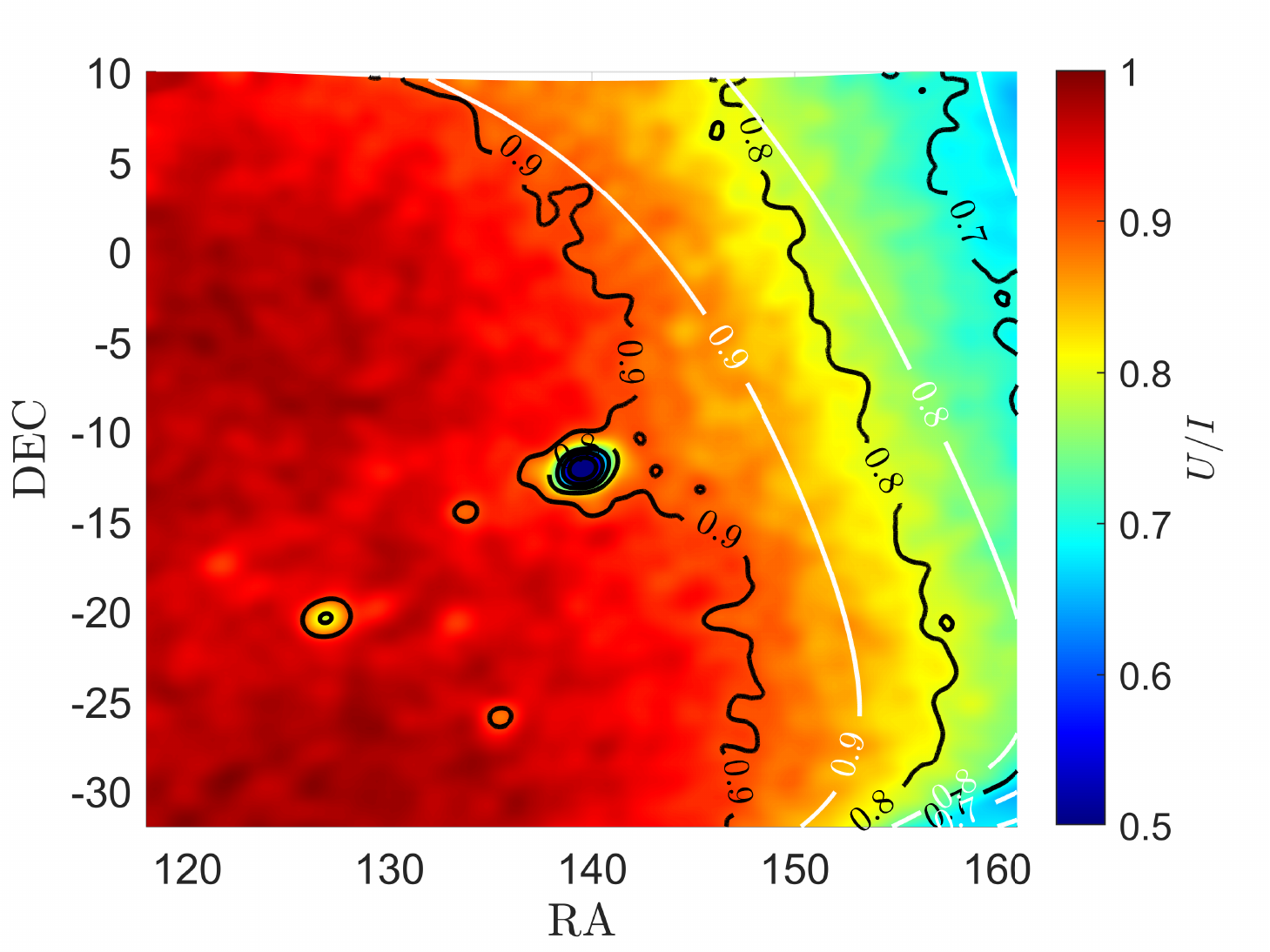}
          \caption{}
         \label{fig:obs_SEFDUI}
     \end{subfigure}
\caption{Side by side comparison of the ratio of the simulated and observed $SEFD_U/SEFD_I$ at gridpoint 114.  (a)~Simulated image of $SEFD_U/SEFD_I$ with the corresponding contour levels. (b)~Observed $SEFD_U/SEFD_I$ with the corresponding contour levels (black curves) and the contour levels of the simulated values (white curves). The observed image has a size of $4096 \times 4096$ pixels and was smoothed using a Gaussian filter of size $201\times201$ to smooth the noisy image.}
\label{fig:stokesUI}
\end{figure*}

 \begin{figure*}[h]
  \centering
  \begin{subfigure}[t]{0.45\textwidth}
         \centering
          \includegraphics[width=\textwidth]{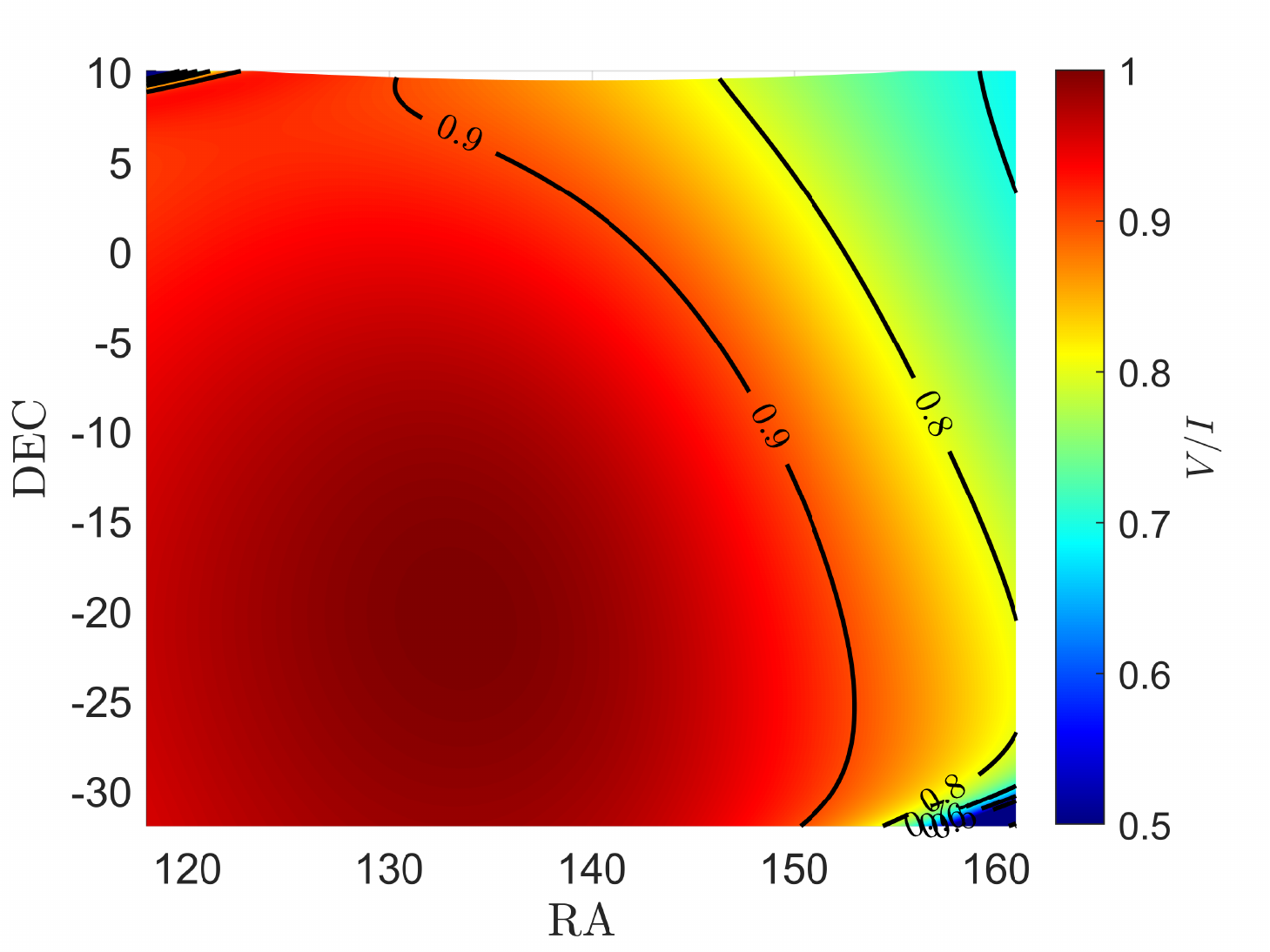}
          \caption{}
         \label{fig:sim_SEFDVI}
     \end{subfigure}
  \hfill
     \begin{subfigure}[t]{0.45\textwidth}
         \centering
          \includegraphics[width=\textwidth]{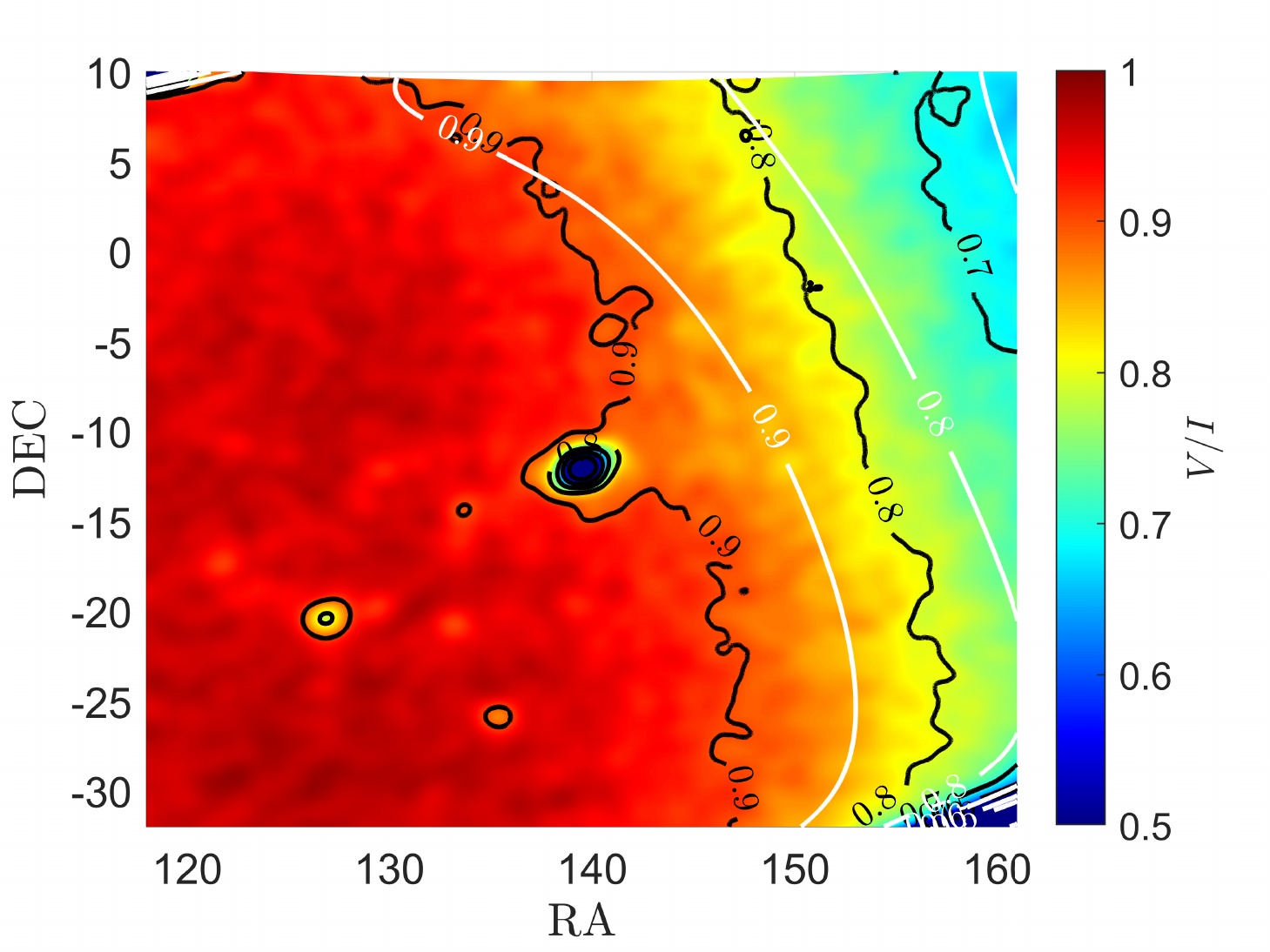}
          \caption{}
         \label{fig:obs_SEFDVI}
     \end{subfigure}
\caption{Side by side comparison of the ratio of the simulated and observed $SEFD_V/SEFD_I$ at gridpoint 114. (a)~Simulated image of $SEFD_V/SEFD_I$ with the corresponding contours levels (black curves). (b)~Observed $SEFD_V/SEFD_I$ with the corresponding contours levels (black curves) and the contour levels of the simulated values (white curves). The observed image has a size of $4096 \times 4096$ pixels and was smoothed using a Gaussian filter of size $201\times201$ to smooth the noisy image.}
\label{fig:stokesVI}
\end{figure*}

\begin{figure*}[h]
  \centering
       \begin{subfigure}[t]{0.45\textwidth}
         \centering
          \includegraphics[width=\textwidth]{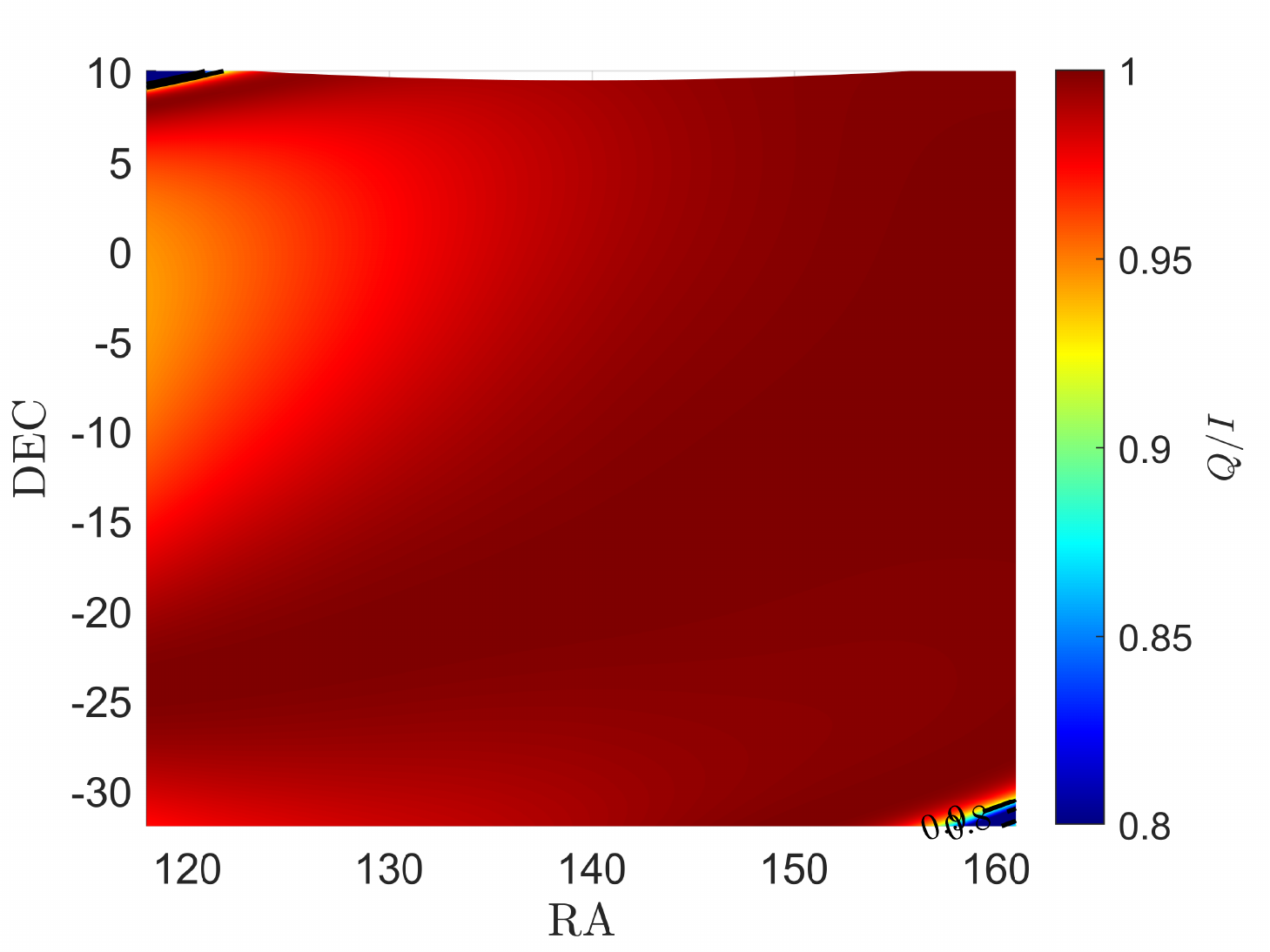}
          \caption{}
         \label{fig:sim_SEFDQI}
     \end{subfigure}
  \hfill
     \begin{subfigure}[t]{0.45\textwidth}
         \centering
          \includegraphics[width=\textwidth]{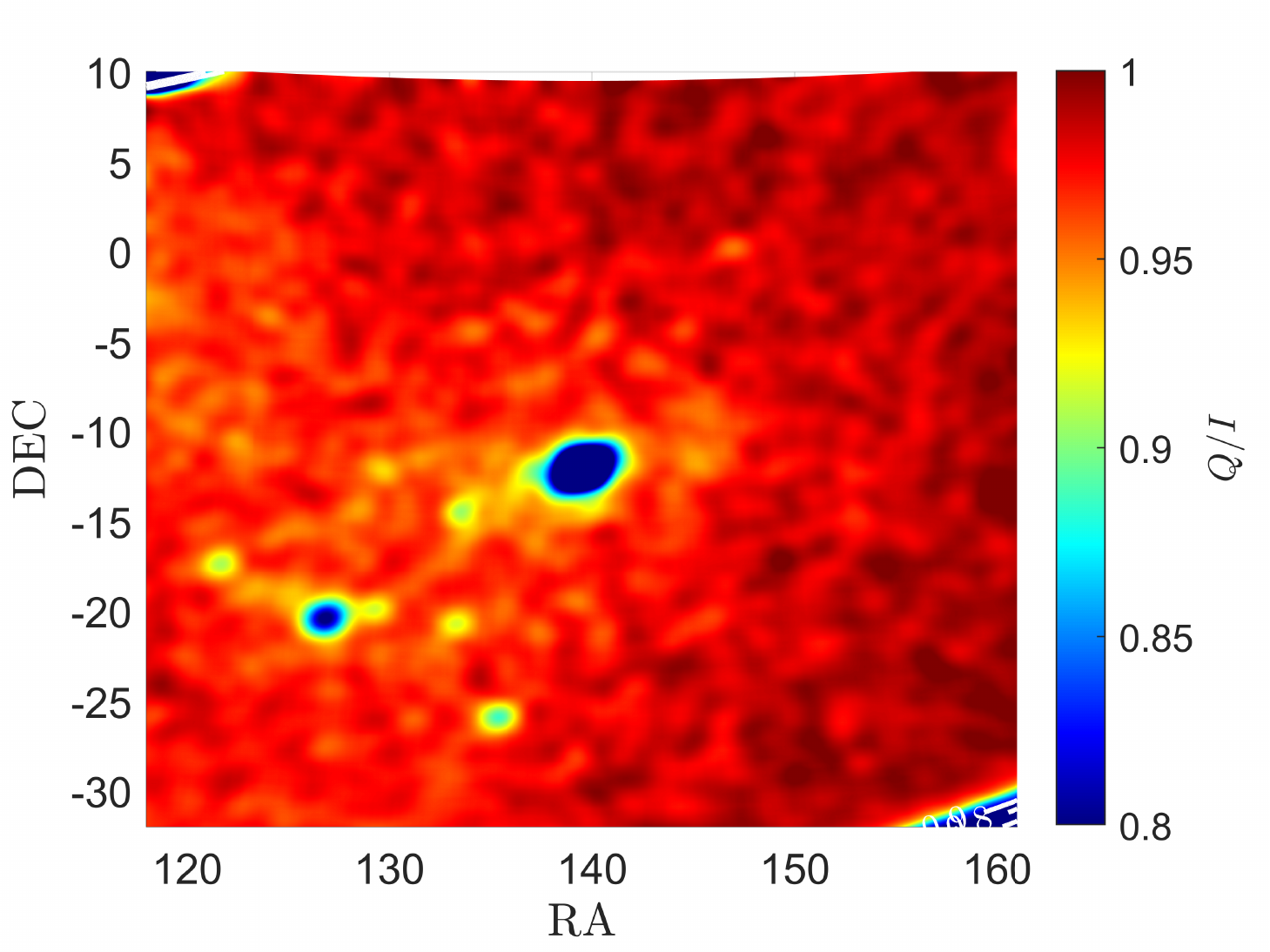}
          \caption{}
         \label{fig:obs_SEFDQI}
     \end{subfigure}
\caption{Side by side comparison of the ratio of the simulated and observed $SEFD_Q/SEFD_I$ at gridpoint 114. (a)~Simulated image of $SEFD_Q/SEFD_I$ with the corresponding contours levels (black curves). (b)~Observed $SEFD_Q/SEFD_I$ with the corresponding contours levels (black curves) and the contour levels of the simulated values (white curves). The observed image has a size of $4096 \times 4096$ pixels and was smoothed using a Gaussian filter of size $201\times201$ to smooth the noisy image.}
\label{fig:stokesQI}
\end{figure*}

The same observational data at $154.88~\si{\mega\hertz}$ as reported in Paper~III was reprocessed and used to compute $\text{SEFD}_{Q, U, V}$ using the same methods as described in Paper~III. For clarity, the observation took place on the 26th December 2014 at 16:05:42 hrs UTC time with the MWA telescope~\citep{tingay_2013} pointing at grid point 114 ($ZA = 46.1\degree,~Az = 80.5\degree$). All observational images have a size of $4096 \times 4096$ pixels and were smoothed in Matlab with a Gaussian filter of size $201\times 201$ to reduce the effects of noise.

The absolute values between simulated and observed $\text{SEFD}_{I}$ were verified in Paper III; therefore, to validate Eq.~\eqref{eqn:SEFD_gen2}, it is sufficient to show that the predicted ratio of $\text{SEFD}_{U}/\text{SEFD}_{I}$, $\text{SEFD}_{V}/\text{SEFD}_{I}$ and $\text{SEFD}_{Q}/\text{SEFD}_{I}$ match observation data. Fig.~\ref{fig:sim_SEFDUI} shows the predicted $\text{SEFD}_{U}/\text{SEFD}_{I}$ calculated using Eq.~\eqref{eqn:SEFD_gen2} based on the simulated MWA array, while Fig.~\ref{fig:obs_SEFDUI} shows the observed $\text{SEFD}_{U}/\text{SEFD}_{I}$. The solid contour lines seen in Fig.~\ref{fig:stokesUI},  Fig.~\ref{fig:stokesVI}, and Fig.~\ref{fig:stokesQI} represent a change in level of 0.1. For ease of comparison, the simulated (black) contour lines shown in Fig.~\ref{fig:sim_SEFDUI} were transferred to Fig.~\ref{fig:obs_SEFDUI} and are shown as white contour lines. The black contour lines in  Fig.~\ref{fig:obs_SEFDUI} represent the contours of the observation data after smoothing. Therefore, the comparison we suggest here is between the white and black contour lines in Fig.~\ref{fig:obs_SEFDUI}. We observe that the simulated contours predict the trends of the observed contours very well. Furthermore, the simulated contours are within a few percent of the observation.

We repeated this process in Fig.~\ref{fig:stokesVI} and Fig.~\ref{fig:stokesQI}. Figs.~\ref{fig:obs_SEFDVI} and~\ref{fig:obs_SEFDQI} show the observed $\text{SEFD}_{V}/\text{SEFD}_{I}$ and $\text{SEFD}_{Q}/\text{SEFD}_{I}$ respectively. The agreement between the simulation and observation in Fig.~\ref{fig:obs_SEFDVI} is very similar to that in Fig.~\ref{fig:obs_SEFDUI}. We calculated average percentage differences between the simulation and observation over the entire images of $\text{SEFD}_{U}/\text{SEFD}_{I}$ and $\text{SEFD}_{V}/\text{SEFD}_{I}$ of approximately $3.6\%$ and $3.5\%$, respectively. Notwithstanding the excellent overall agreement over the images, we see a small systematic offset of approximately 3.5\% between the observed and simulated contour lines in Figs.~\ref{fig:obs_SEFDUI} and~\ref{fig:obs_SEFDVI}. The exact reason for this is not fully understood. However, we note that agreement of the $SEFD_U/SEFD_I$ and $SEFD_V/SEFD_I$ is similar to the level of agreement within a few percent between simulation and observation in our previous work~\citep{Sutinjo_AA2021, Sutinjo_Array_AnA} and is expected. A further contributing factor to the offset could be the presence of a non-functional dipole in an MWA tile due to a damaged low noise amplifier. Preliminary simulations of an MWA tile with a non-functional dipole suggest the systematic offset of a few percent could vanish or increase depending on the location of the broken dipole. However, the inclusion of models which include random non-functioning dipoles throughout the entire MWA telescope requires extensive further study which may be addressed in future work.


In Fig.~\ref{fig:obs_SEFDQI}, we note that while contours lines of the simulated values are included, they are not visible as they only start to change from 1 to 0.9 at the top left and bottom right corners. 

 We see an excellent agreement between the simulated and observed $\text{SEFD}_{Q}/\text{SEFD}_{I}$ with an average percentage difference of $1.8\%$ between the observed and simulated results. The near equality between $\text{SEFD}_I$ and $\text{SEFD}_Q$ seen in the image is consistent with that suggested by the idealized calculation in Sect.~\ref{sec:short_dip}.  The noticeable object in the center of Figs.~\ref{fig:obs_SEFDUI},~\ref{fig:obs_SEFDVI}, and \ref{fig:obs_SEFDQI} is the calibrator source Hydra-A which was the main target of the analyzed observation. 
 This is an artifact of the difference imaging and is not relevant to SEFD calculation.

The simplified calculation in Sect.~\ref{sec:short_dip} suggests that $\text{SEFD}_U/\text{SEFD}_I$ and $\text{SEFD}_V/\text{SEFD}_I$ are less than or equal to unity for X-Y dipoles. This is consistent with Figs.~\ref{fig:obs_SEFDUI} and~\ref{fig:obs_SEFDVI}.  In addition, it is also evident by comparing the contour shapes in Fig.~\ref{fig:obs_SEFDUI} and \ref{fig:obs_SEFDVI} that $\text{SEFD}_U$ and $\text{SEFD}_V$ are nearly equal, which is again consistent with Sect.~\ref{sec:short_dip}. These agreements are expected because the MWA antennas at this observing frequency can be approximated as X-Y short dipoles.

\section{Scientific applications}
\label{sec_science_apps}

Although the majority of radio sources are unpolarized, some astrophysical objects of transient nature, such as pulsars, fast radio bursts (FRBs), M dwarf stares, exhibit large degree of polarization. For example, \citet{2018MNRAS.478.2835L} analyzed MWA data at 169 -- 200\,MHz band and demonstrated detection of 33 pulsars in Stokes V images. Similarly, at GHz frequencies, \citet{2019ApJ...884...96K} serendipitously found an new pulsar PSR J1431-6328 as a highly polarized steep-spectrum point source in a deep image from the Australian Square Kilometre Array Pathfinder telescope \citep[ASKAP;][]{2014PASA...31...41H,2016PASA...33...42M}. More recently, \citet{2022arXiv220500622W} also analyzed ASKAP data and discovered another highly circularly polarized ($\sim$20\%), variable, steep-spectrum pulsar PSR J0523-7125 in the Large Magellanic Cloud (LMC), while \citet{2021ApJ...920...45W} found a highly polarized ($\sim$80\% linear and 10 -- 20\% circular polarization) transient ASKAP J173608-2321635 of yet unknown nature. In the recent search in ASKAP data, \citet{2021MNRAS.502.5438P} reported detection of 33 circularly polarized stars of variuous types (including M dwarfs), and 23 of them did not have earlier detections in radio. Further, \citet{2019ApJ...871..214V} observed 5 nearby M dwarf stars and detected 22 flares with high degree of circular polarization (40 -- 100\%). Another search with ASKAP led to detections of coherent, type IV, radio bursts accompanied by an optical flare from a nearby star Proxima Centauri \citep{2020ApJ...905...23Z}, while a few years earlier \citet{2019MNRAS.488..559Z} detected several coherent highly polarized outbursts from M dwarf star UV Ceti in ASKAP images. Similar observations of UV Ceti with the MWA by \citet{2017ApJ...836L..30L} led to detection of four flares in polarization images with circular and linear polarization fractions of $>$27\% and $>$18\% respectively. All these results demonstrate that imaging wide fields of view (FoV) in all Stokes polarizations opens a very promising avenue for finding new pulsar candidates, outbursts of M dwarf stars and potentially other transient phenomena.

Furthermore, FRBs are one of the most exciting puzzles of the high-energy astrophysics, and their polarization measurements can provide crucial input for understanding progenitors and physical mechanisms behind these enigmatic events \citep{2021ApJ...920...46D}. They enable characterization of magneto-ionic properties of the interstellar medium (ISM) and immediate local environments of FRBs \citep[e.g.][]{2022Sci...375.1266F,2022MNRAS.511.6033P}. Furthermore, they can ultimately help explaining FRB emission mechanisms as various theoretical models predict different fractions of circular polarization and behaviours of position angles \citep[e.g.][]{2022arXiv220205475T,2021arXiv211206719W}. The leading two models for FRB emission invoke either magnetospheric origin of coherent emission or synchrontron maser mechanism in relativistic shocks from a highly magnetized neutron star (NS), and \citet{2021ApJ...920...46D} argues that detection of circular polarization will distinguish between these different models. High degree of polarization have been reported in one-off FRBs \citep[e.g.][]{2019MNRAS.486.3636P}, and repeating FRBs \citep{2022MNRAS.512.3400K}. In particular, the repeating FRBs, which can be regularly monitored and re-detected, offer opportunities to fully characterize their emission properties, including polarization. For example, recent observation of polarization position angle (PA) swings in 7 out of 15 bursts from repeating FRB 180301 \citep{2020Natur.586..693L,2019MNRAS.486.3636P}, and high degree of linear polarization observed in FRB 20180916B at 1.7 GHz \citep{2021NatAs...5..594N} both support magnetospheric origin of the emission. However, more observational data are required to confirm the physical mechanisms powering these extreme events.

\section{Conclusions}
\label{sec_conclusions}
As discussed in Section~\ref{sec_science_apps}, a wealth of information about astrophysical objects, of transient nature in particular, can be derived from sensitive polarization measurements. Therefore, in order to unlock full science potential of the existing and future low frequency instruments, such as the MWA and SKA-Low respectively, we need to understand and characterize sensitivity in wide FoV images in all Stokes polarizations.

Building upon Paper~II and~III, we presented a general formula to calculate the $\text{SEFD}_{I,Q,U,V}$ of polarimetric interferometers. The key result for dual-polarized antennas, which is most relevant to ground-based radio astronomy, is Eq.~\eqref{eqn:SEFD_gen2}. This equation is to be used with the corresponding Eq.~\eqref{eqn:MIdual} (Stokes~I), Eq.~\eqref{eqn:MQdual} (Stokes~Q), Eq.~\eqref{eqn:MUdual} (Stokes~U), and Eq.~\eqref{eqn:MVdual} (Stokes~V). We verified this equation by re-processing the MWA observation dataset used in Paper~III to obtain $\text{SEFD}_{Q,U,V}$. We then took the ratios $\text{SEFD}_Q/\text{SEFD}_I$, $\text{SEFD}_U/\text{SEFD}_I$, and $\text{SEFD}_V/\text{SEFD}_I$. We showed that the observed ratios match the predicted results calculated using Eq.~\eqref{eqn:SEFD_gen2} with the average percentage difference of $1.8\%$, $3.6\%$ and $3.5\%$ for $\text{SEFD}_Q/\text{SEFD}_I$, $\text{SEFD}_U/\text{SEFD}_I$ and $\text{SEFD}_V/\text{SEFD}_I$ respectively. These results suggest that the formulas derived in this paper are valid. 

In addition, we presented $\text{SEFD}_{I, Q, U, V}$ formulas in Eq.~\eqref{eqn:SEFD_I_Q} and Eq.~\eqref{eqn:SEFD_U_V} for the special case of  X-Y crossed short dipoles which are often used in low-frequency radio astronomy. These results show that the $\text{SEFD}_{I, Q, U, V}$ are only dependent on the zenith angle and independent of the azimuth angle. Furthermore for this special case, $\text{SEFD}_I$ is equal to $\text{SEFD}_Q$ and $\text{SEFD}_U$ is equal to $\text{SEFD}_V$; $\text{SEFD}_U$ and $\text{SEFD}_V$ are less than or equal to $\text{SEFD}_I$ and $\text{SEFD}_Q$. 

\bibliographystyle{aa}
\bibliography{Final_SEFD_IQUV.bib}

\end{document}